# Using Models at Runtime to Address Assurance for Self-Adaptive Systems

Betty H.C. Cheng[1], Kerstin I. Eder[2], Martin Gogolla[3], Lars Grunske[4], Marin Litoiu[5], Hausi A. Müller[6], Patrizio Pelliccione[7], Anna Perini[8], Nauman A. Qureshi[9], Bernhard Rumpe[10], Daniel Schneider[11], Frank Trollmann[12], and Norha M. Villegas[6,13]

[1] Michigan State University, US
chengb@cse.msu.edu
[2] University of Bristol, UK
Kerstin.Eder@bristol.ac.uk
[3] Universität Bremen, Germany
gogolla@informatik.uni-bremen.de
[4] TU Kaiserslautern, Germany
grunske@informatik.uni-kl.de
[5] York University, Canada
mlitoiu@yorku.ca
[6] University of Victoria, Canada
hausi@cs.uvic.ca
[7] Università degli Studi dell'Aquila, Italy
patrizio.pelliccione@univaq.it,
Chalmers University of Technology and University of Gothenburg, Sweden
patrizio.pelliccione@gu.se
[8] CIT - FBK - Povo Trento, Italy
perini@fbk.eu
[9] National University of Sciences and Technology (NUST), Pakistan
nauman.qureshi@seecs.edu.pk
[10] RWTH Aachen, Germany
rumpe@se-rwth.de
[11] Fraunhofer IESE - Kaiserslautern, Germany
daniel.schneider@iese.fraunhofer.de
[12] TU Berlin, Germany
Frank.Trollmann@dai-labor.de
[13] Icesi University, Colombia
nvillega@icesi.edu

**Abstract.** A self-adaptive software system modifies its behavior at runtime in response to changes within the system or in its execution environment. The fulfillment of the system requirements needs to be guaranteed even in the presence of adverse conditions and adaptations. Thus, a key challenge for self-adaptive software systems is assurance. Traditionally, confidence in the correctness of a system is gained through a variety of activities and processes performed at development time, such as design analysis and testing. In the presence of self-adaptation, however, some of the assurance tasks may need to be performed at runtime. This need calls for the development of techniques that enable continuous assurance throughout the software life cycle. Fundamental to the development of runtime assurance techniques is research into the use of models at runtime

(M@RT). This chapter explores the state of the art for using M@RT to address the assurance of self-adaptive software systems. It defines what information can be captured by M@RT, specifically for the purpose of assurance, and puts this definition into the context of existing work. We then outline key research challenges for assurance at runtime and characterize assurance methods. The chapter concludes with an exploration of selected application areas where M@RT could provide significant benefits beyond existing assurance techniques for adaptive systems.

## 1 Introduction

A self-adaptive system (SAS) modifies its behavior at runtime in response to changes in the system itself or in its environment.[1] An SAS generally comprises a component that delivers the basic function or service, often referred to as the *target* or *managed system*, and another component that controls or manages that target system through an *adaptation process*, often referred to as the *controller* [MAB$^+$02] or *autonomic manager* [KC03]. The target system can be viewed as a steady-state program [ZC06a, GCZ08]. It is not adaptive and is applicable to a specific execution environment. The SAS controller can, via the invocation of an adaptation process that implements *adaptive logic* [ZC06a], transform this steady-state program to a different steady-state program—one that is suitable for a different set of environmental conditions [ZC06a]. As such, the steady-state program that delivers the basic function or service of an SAS is the target of the adaptation process that is managed by the controller. During the adaptation process, it is important to provide assurance that the system does not become inconsistent (e.g., no data is lost and transactions are not interrupted) [KM90, ZCYM05, ZC06b].

The IEEE Standard Glossary of Software Engineering Terminology defines *assurance* as "a planned and systematic pattern of all actions necessary to provide adequate confidence that an item or product conforms to established technical requirements" [IEE90].[2] For non-adaptive systems, assurance is typically performed at design and development time. In practice, assurance tasks comprise verification, validation, test, measurement, conformance to standards, and certification. Collectively, these tasks all contribute to gaining confidence that both the processes employed and the end product satisfy established technical requirements, standards, and procedures. In the presence of runtime adaptations in an SAS, the fulfillment of the system requirements need to be guaranteed at runtime, even during the adaptation process [ZC05, ZC06b, VMT$^+$11b]. Thus, software assurance becomes a critical runtime concern, giving rise to the need for continuous assurance over the entire life cycle of a software system. Given the increasing use of SASs in safety-critical applications (e.g., power-grid management, transportation management systems, telecommunication systems, and health-monitoring), assurance for SASs is of paramount importance. The development of rigorous methods and techniques that extend assurance from development time to runtime is therefore a high priority on the research agenda for the SAS research community.

---

[1] This chapter uses the acronym SAS to refer to any software-based system that exposes self-* features.

[2] This chapter uses the term *software assurance* rather than the more specific term *software quality assurance* to not only include software quality concerns but also safety, reliability, and security concerns.

Assurance is required for both functional properties (i.e., those describing specific functions of the system such as the result of a computation) and non-functional properties (i.e., those describing the operational qualities of the system such as availability, efficiency, performance, reliability, robustness, security, stability, and usability) [VMT[+]11b]. Guaranteeing these properties at runtime in SASs is particularly challenging due to the varying assurance needs posed by a changing system or execution environment, both fraught with uncertainty [RJC12, EM13]. Nevertheless, the properties specified in the system requirements need to hold *before*, *during*, and *after* adaptation [ZC06a, ZC06b, ZGC09].

Continuous assurance throughout the entire software life cycle provides unprecedented opportunities for monitoring, analyzing, guaranteeing, and predicting system properties throughout the operation of a software system. The fact that many variables that are free at development time are bound at runtime enables us to *tame* the state space explosion, thus enabling the exploration of states that could not have been considered at development time. This reduction in state space provides new opportunities for runtime verification and validation (V&V), leading to assurance of critical system properties at runtime [TVM[+]12]. Fundamental to the development of runtime assurance techniques is research into models that can be used at runtime.

This chapter presents models at runtime (M@RT) as a foundation for the assurance of SASs and discusses related research challenges. Section 2 reviews assurance criteria, both functional and non-functional, whose fulfillment depends on or can be affected by self-adaptation and therefore requires assurance at runtime. Section 3 classifies different types of models used for M@RT and discusses the application of M@RT to support a spectrum of assurance issues. Section 4 identifies research challenges in the area of M@RT for SAS assurance tasks. Section 5 characterizes existing methods used for assurance of SASs. Section 6 describes selected application areas that exhibit the type of assurance challenges that we consider amenable to the use of M@RT. Finally, Section 7 concludes the chapter.

## 2   Assurance Criteria for Self-Adaptive Software Systems

Assurance criteria for SASs include functional and non-functional requirements whose fulfillment depends on or can be affected by self-adaptation. It is important to distinguish between assurance criteria applicable to the *target system* (i.e., criteria that relate to properties of the current or a potential future state of that system), and assurance criteria applicable to the *adaptation process* itself. Sections 2.1 and 2.2 respectively discuss functional and non-functional requirements as fundamental assurance criteria for SASs.

### 2.1   Functional Requirements

A functional requirement specifies a function that a system or system component must be able to perform [IEE90]. Functional requirements are typically formulated as prescriptive statements to be satisfied by the system. While it is still a common practice to describe functional requirements using natural language, the potential for misinterpretation of such descriptions is considerable due to the inherent ambiguity of natural languages [Ber08, CNdRW06]. Formal languages with well-defined semantics provide a more rigorous and reliable means for specifying functional requirements in the context of system design. The following discussion is limited to formal descriptions.

Functional requirements describe the behavioral objectives of the functions $f$ of a system. They are typically defined in terms of relating the inputs $I$ to the system with the outputs $O$ of the system, with the expectation that $f : I \rightarrow O$. A function $f$ may be some type of computation, data manipulation, or other specific functions that the system should execute. Accordingly, the input $I$ may be data from a user, values from a sensor, such as a temperature value or a sequence of images. Similarly, the output $O$ may be pictures, continuous video, a braking signal for a car, or the opening of a valve. It is important ot note that functional requirements describe the system behavior that is visible at the system boundaries (i.e., system interfaces) [ZJ97]. The boundaries can be at the human-computer interface, sensors, actuators, or even at the boundaries between interacting systems. As such, functional requirements describe "what" the system has to provide in terms of its functional behavior to meet the expectations of its users, leaving "how" this functionality will be achieved to the design and implementation of the system.

System adaptation may become necessary to handle changes in the requirements or in the environment that are visible at its boundaries and influence its behavior externally. These adaptations may lead to internal changes that manifest as changed behavior observable at the system boundary. While the former is a reaction to the system context and leads to retaining the functional behavior in the presence of external change, the latter is a reaction to changing user needs or system configuration needs and leads to behavioral adaptations to accommodate the new requirements.

Because an SAS tends to respond to changes in the environment, functional requirements should take into account the context of the system as well as explicit assumptions about its behavior. Adaptation provides a means to alter the way a system satisfies its functional requirements, including the use of machine learning techniques [KM07], agent-based techniques [SAS14], bio-inspired techniques [BSG$^+$09, MV14], and selecting specific target configuration from a collection of different target configurations [GCH$^+$04, ZC06a], each of which satisfies the functional requirements, but may be better suited for a specific context and/or set of environmental conditions. The functional requirements may be formalized in an "assume/guarantee" style [JT96]—assuming a set of conditions or restrictions holds, then the application of the function guarantees that the results satisfy a set of required properties. The definition of pre- and postconditions is an example of this style of functional requirements specification.

Common formalisms used to express functional requirements are Linear-Time Temporal Logic (LTL) [Pnu81] and Computational Tree Logic (CTL) [BAMP81], both of which are included in the logic CTL* [CE82]. Several languages have been proposed to facilitate the specification of functional properties; examples range from basic assertion languages such as PSL [Acc04], used in electronic system design, to scenario-based visual languages, such as Message Sequence Charts [HT04] or Property Sequence Charts [AIP07]. These languages are often less expressive than pure temporal logic, but are designed to be intuitive and user friendly.

Beyond property-based specification, various algebraic specification and system modeling techniques have been developed, including Statecharts [Har87]; set-theoretic approaches, such as VDM [BJ78] and Z [ASM80]; process or operational-oriented, including SDL [Uni99], the B Method [Abr88], Event-B[ABH+10]; object-oriented languages, such as UML and its numerous variants[3]; architectural description languages [Cle96]; and Matlab/Simulink[4] to name a few representative examples. Traditionally, these techniques are used during system design and development to achieve increased confidence in the functional correctness of the system. Several of the above listed techniques support automatic code generation from the system model as well as formal verification at varying levels of abstraction.

Several complementary approaches have been used to specify functional requirements of an SAS, where uncertainty of the execution environment is implicitly or explicitly acknowledged by allowing more flexibility in how requirements can be satisfied. The SAS determines at runtime how to realize the specified functionality when placed in its target environment. This flexibility can be achieved by describing functional requirements in terms of policies that encode high-level specifications of functional objectives together with a set of operational constraints. This implicit approach to acknowledging uncertainty in the execution environment can utilize utility functions and a rule-based approach in the context of a goal-oriented functional requirements specification. Another approach is to explicitly acknowledge specific system functionality affected by uncertainty and thus allow specific points of flexibility in satisfying the requirements, such as that provided by the RELAX [WSB+09, CSBW09, RFJB12, FDC14a] and FLAGS [BPS10, PS11] approaches. Section 5.1 provides further details on these approaches.

### 2.2 Non-functional Requirements

If we consider functional requirements of a software system to be a function $f$ that directly maps input $I$ to output $O$ ($f : I \rightarrow O$), then non-functional requirements refer to properties about $f$, $I$, $O$ or relationships between $I$ and $O$ [CPL09]. Non-functional requirements such as performance, dependability, safety, security, and their corresponding quality attributes such as latency, throughput, capacity, confidentiality, and integrity can include assurance concerns from the perspective of both the target system and the adaptation mechanism. Avižienis *et al.* [ALRL04] and Barbacci *et al.* [BKLW95] provide two comprehensive taxonomies of software quality attributes useful for the identification of assurance criteria in SASs.

It is necessary to validate and continually monitor non-functional requirements on both the target system and the adaptation process using techniques such as probabilistic monitoring [GZ09, Gru11], requirements monitoring [FF95], [FFvLP98], or utility function monitoring [GCH+04, RC11]. At runtime, the desired properties of the target system may no longer hold due to changes in the target system's context of use (e.g., user, platform, or environment context [SCF+06]), or side effects introduced by adaptations. In the latter case, it is possible to derive the impact of adaptations on properties of the target system by analyzing adaptation properties such as stability, accuracy, settling time, small overshoot, and robustness. Specifically, it may be possible to take advantage of this relation to detect consequences of adaptations performed by controllers [KC03] or consequences of a changing environment (e.g., a failing component or a deficient Internet connection).

---

[3] www.uml.org
[4] http://www.mathworks.com

Several non-functional assurance criteria may be more easily guaranteed at runtime than at design time. For example, it is easier to assess latency when it is possible to measure and continually monitor delay times in the running system. Table 1 presents examples of non-functional assurance criteria with corresponding quality attributes (cf. Columns 1 and 2). Adaptation properties (cf. Column 3), defined as assurance criteria that concern the adaptation process [VMT[+]11b], can be mapped to quality attributes measurable at runtime for both the target system and the adaptation mechanism. Where to measure a given property, either in the adaptation process or in the target system, will depend on its definition and its assessment metric. For example, *settling time* defined as the time required for the adaptation process to take the target system to a desirable state, must be measured on the target system since the need for the adaptation and the conditions for a desired state can only be observed at this level. Moreover, settling time can be measured through different quality attributes, depending on the specific non-functional property that must be satisfied. For example, if the concern is performance, settling time can be observed in terms of the time the system takes to perform a particular process. When the accepted time limit for this process is exceeded, the adaptation process will be invoked. Once the process execution time is back within desired limits, the target system will have reached its desired state. As such, settling time is the time elapsed between the moment at which the need for adaptation was detected and the moment at which the system reaches the desired new state. Villegas *et al.* [VMT[+]11b] provide a comprehensive catalogue of adaptation properties and the corresponding quality attributes needed to identify the assurance criteria applicable to the adaptation process. This study also surveys definitions for the assurance criteria presented in Table 1.

**Table 1.** Examples of non-functional assurance criteria that are better guaranteed at runtime than at design time (including their mapping to quality attributes and adaptation properties) [VMT[+]11b]

| Assurance Criteria | Quality Attribute | Adaptation Properties |
|---|---|---|
| Latency | Performance | Stability, accuracy, settling time, overshoot, scalability |
| Throughput | Performance | Stability, accuracy, settling time, overshoot, scalability |
| Capacity | Performance | Stability, accuracy, settling time, overshoot, scalability |
| Safety | Dependability | Stability |
| Availability | Dependability | Robustness, settling time |
| Reliability | Dependability | Robustness |
| Confidentiality | Security | Security |

Assuring these criteria at runtime requires effective monitoring mechanisms and M@RT to analyze, guarantee, and predict the qualities of the target system and the adaptation process dynamically. Implementing these mechanisms effectively requires a thorough analysis of the interdependencies between non-functional assurance criteria, quality attributes, and adaptation properties as presented in Table 1. This mapping constitutes a valuable starting point to identify assurance criteria and adaptation properties. On the one hand, this mapping supports the identification of assurance criteria according to the target system's desired quality attributes. (For example, latency, throughput and capacity are relevant assurance criteria when performance is the negotiated quality attribute.) On the other hand, it is useful to identify adaptation properties, relevant to quality attributes, that are applicable to the adaptation mechanism. (For example, when performance is a key quality attribute for the target system, then stability, accuracy, settling time, small overshoot, and scalability constitute relevant properties to be guaranteed in the adaptation process.) Of course these mappings also depend on the actual target system, its technical implementation, and the performed adaptations.

## 3 Models at Runtime

SASs require rethinking the notion of the software life cycle for which the distinction between development time and execution time stages is no longer starkly apparent (e.g., PLASTIC,[5], SMScom[6]). Recent approaches recognize the need to produce, manage, and maintain software models all along the software's life time to assist the realization and validation of system adaptations while the system executes [Inv07, BBF09, BG10, ACR$^+$11, BDM$^+$11, VTM$^+$12, MV14] [CVM14].

Continuing with this line of reasoning, our objective is to explore models of different aspects of the application (e.g., requirements, specification, design, architecture, implementation, infrastructure, instrumentation, and context-of-use) and life cycle phases (e.g., design time, development time, configuration time, load time, and runtime) to deal with the inherent dynamics of self-adaptation in software systems. These abstractions, combined with suitable instrumentation, could provide effective techniques for monitoring, analyzing, guaranteeing, and predicting system properties throughout the operation of an SAS.

The kind of models used at runtime can be classified by (1) their purpose—predictive, prescriptive, constructive, or descriptive; (2) their underlying modeling languages—for example, the 14 UML 2.2 structural and behavioral diagrams, State-charts, Petri Nets, and logic based models (e.g., Temporal Logics); and (3) the aspects they describe—data structure, task or process state, I/O behavior, or interaction pattern.

One of the main principles of using M@RT for assurance is to exploit the causal connection [Mae87] between the model and the system under development at runtime. This connection determines synchronization between the model and the running system. For example, M@RT can be updated to reflect changes in the running system —we say that they are in *descriptive causal connection*. This type of modeling enables assurance techniques to analyze abstract models instead of the actual implementation of the application when collecting information for assurance. In contrast, the model can be changed to cause an adaptation of the application (i.e., *prescriptive causal connection*). This use of modeling can be used to implement adaptations of the running system that are required to assure system properties.

---

[5] FP6 IST EU PLASTIC project http://www.ist-plastic.org
[6] Carlo Ghezzi, Self-Managing Situated Computing Grant, ERC Advanced Investigator Grant N. 227977, European Union, 2008–2013

In the scope of assurance, M@RT can be used as a basis for assuring functional as well as non-functional properties of the system (cf. Section 2). From this perspective, models can play various roles. Depending on what the models describe, they can be used as a source of information about aspects of the running system. For instance, goal models can represent the requirements that need to be assured, the current state of the system, adaptations, or the context of use. M@RT can have several purposes for runtime assurance. Among others, they can be used as information sources for monitoring aspects of a running system, to influence the system via model manipulation, and as a basis for analysis methods, such as model-based verification and model-based simulation. For analysis methods, models are usually beneficial as they provide easy to use high-level knowledge about the system.

Development-time modeling approaches already exploit these advantages and enable the assertion of certain properties of a developed system. The use of M@RT has the advantage that some of the analysis constraints are relaxed as the current runtime state is available for reasoning, reaction, and regulation. At development time, full assurance is required to reason about all possible states. Several of these variables that are unknown at development time are bound at runtime and can allow for a more focused analysis of the current state and possibly several neighboring ones. This variable instantiation is especially useful for factors that can only be estimated at development time (e.g., network delay). A running system can continually monitor these aspects and react to them. The remainder of this section describes the dynamics of adaptive systems and the use of models during the adaptation process.

### 3.1  M@RT and the Dynamics of Self-Adaptive Software

The Software Engineering for Adaptive and Self-Managing Systems (SEAMS) community has identified three key subsystems needed for the design of effective context-driven self-adaptation: the control objectives manager, the adaptation controller, and the context monitoring system [VTM$^+$12]. These subsystems represent three levels of dynamics in self-adaptation, each of which can be controlled through a corresponding feedback loop. Villegas *et al.* [VTM$^+$12] provide a comprehensive characterization of these three levels of dynamics in SASs.

In general, assurance criteria drive the control objectives, adaptation, and monitoring feedback loops, as well as their interactions. As such, assurance governs the behavior of both the target system and the adaptation process. For example, system administrators can provide the control objectives manager with the required specifications. More specifically, the control objectives manager then sends the adaptation goals to the adaptation controller and monitoring requirements to the monitoring system. Thus, these specifications govern the behavior of the adaptation process and the behavior of the SAS throughout the adaptation process.

We argue that M@RT provide abstractions that are essential to support the feedback loops that control the three levels of dynamics identified in SASs. From this perspective, M@RT (cf. Figure 1) could be developed specifically for each level of dynamics to support the control objectives manager, adaptation controller, and the monitoring system. The figure also shows the interactions between these models and the respective subsystems in an SAS.

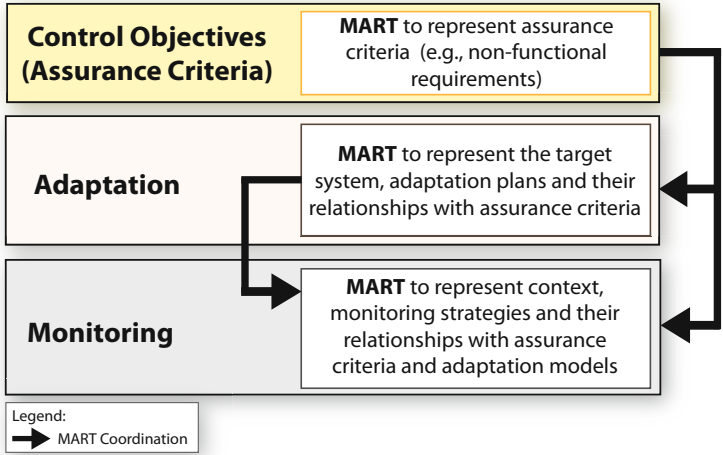

**Fig. 1.** The three levels of M@RT for the assurance of SASs

- At the *Control Objectives* level, M@RT represent requirements specifications subject to assurance in the form of functional and non-functional requirements.
- At the *Adaptation* level, M@RT represent states of the managed system, adaptation plans and their relationships with the assurance specifications.
- At the *Monitoring* level, M@RT represent context entities, monitoring requirements, as well as monitoring strategies and their relationships with assurance criteria and adaptation models.

Most importantly, M@RT at these levels must have efficient and effective methods of inter-level interaction since changes in requirement specifications may trigger changes at both the adaptation and the monitoring levels, as well as in the associated runtime models. Similarly, changes in adaptation models may imply changes in monitoring strategies or context entity models. In any case, M@RT at the adaptation and monitoring levels must maintain an explicit mapping to the models defined at the control objectives level that specify the requirements.

In summary, the architecture of SASs contains three interacting but functionally self-contained levels, each dedicated respectively to control objectives, adaptation, and monitoring of the SAS. Designing an SAS *for assurance*, as opposed to leaving assurance until after system design, requires the tight integration of assurance objectives into each level in the SAS architecture. We argue that this integration can most effectively be achieved by introducing dedicated M@RT that embody specific assurance criteria, focused either for the target system or the adaptation process.

### 3.2 Models at Runtime during the Adaptation Process

As a starting point for a research methodology we analyzed the MAPE-K loop in further detail. Kephart and Chess proposed this autonomic manager as a foundational component of IBM's autonomic computing initiative [KC03]. It constitutes a reference model for designing and implementing adaptation mechanisms in SASs. The MAPE-K loop is an abstraction of a feedback loop where the dynamic behavior of a managed system is controlled using an autonomic manager. The MAPE-K comprises four phases—Monitor (M), Analyzer (A), Planner (P) and Executor (E)—that operate over a knowledge base (K). Each of these phases is briefly described next.

1. *Monitors* gather and pre-process relevant context information from entities in the execution environment that can affect the desired properties and from the target system;
2. *Analyzers* support decision making on the necessity of self-adaptation;
3. *Planners* generate suitable actions to affect the target system according to the supported adaptation mechanisms and the results of the Analyzer;
4. *Executors* implement actions with the goal of adapting the target system; and
5. A *Knowledge Base* enables data sharing, data persistence, decision making, and communication among the components of the feedback loop, as well as arrangements of multiple feedback loops (e.g., the Autonomic Computing Reference Architecture (ACRA) [IBM06]).

In order to illustrate the role of M@RT as enablers of assurance mechanisms for self-adaptation, Figure 2 presents an extension of the MAPE-K loop, where assurance tasks complement each stage of the loop [TVM$^+$12], and the knowledge base is replaced by M@RT. We aptly name the feedback loop depicted in this figure *MAPE-MART loop*.

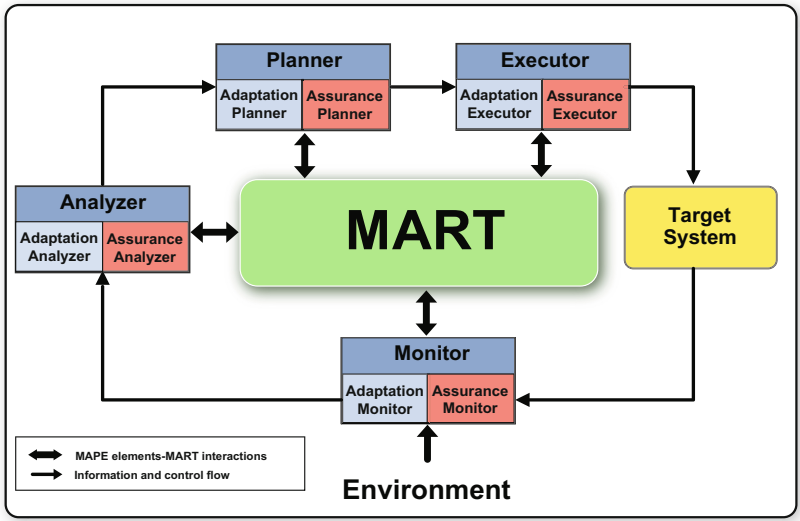

**Fig. 2.** MAPE-MART loop: The MAPE-K loop from autonomic computing extended with M@RT, and assurance instrumentation as foundational elements for the assessment of SASs

MAPE elements interact with M@RT along the adaptation process to either obtain or update information about system states, the environment, and assurance criteria. *Monitors* keep track of relevant context information according to monitoring conditions in the system itself (*assurance monitors*) and its adaptations (*adaptation monitors*). For example, monitors interact with M@RT in order to make monitored data available throughout the adaptation process, or to monitor the states of models or changes in assurance criteria. *Analyzers* will then use monitored context to identify whether desired conditions are being or could potentially be violated. Analyzers can also update models with identified symptoms. Again, we can distinguish between *assurance analyzers* that analyze the system and *adaptation analyzers* that analyze the adaptation process. *Adaptation planners* use the symptoms provided by analyzers to define a new adaptation plan. Adaptation plans can be defined in the form of models that are processable by executors to adapt the target system. Then *assurance planners* check whether the plan is correct with respect to the assurance criteria. Finally, *adaptation executors* perform the plan, after which point, *assurance executors* check whether both the system remains in a safe state and the desired properties are achieved. These verification tasks can be optimized using M@RT.

## 4 Research Challenges for Assurance at Runtime

This section overviews selected research avenues and research challenges for the assurance of SASs using M@RT.

### 4.1 Research Avenues

Software assurance is a large field with many subfields (e.g., software quality, V&V, safety, trust, and several 'ilities') that spans the realms of software engineering, systems engineering, control engineering, and many other engineering disciplines. From a software engineering perspective, assurance at *runtime* for SASs appears to be an emerging area of research [GCZ08, FDB+08, IPT09, TVM+12, FGT11, SBT11, FRC13a, FDC14b]. In contrast, runtime assurance in control engineering traces its roots to the industrial revolution, applied to devices such as the centrifugal governor. This device used a flyball mechanism to sense the rotational speed of a steam turbine and to adjust the flow of steam into the machine. By regulating the turbine's speed, it provided the safe, reliable, and consistent operation that enabled the proliferation of steam-powered factories [MAB+02].

In an instrumented, interconnected, and intelligent world, control and runtime assurance are core components in SASs, providing high performance, high confidence, and reconfigurable operation in the presence of uncertainties. The continuous integration of sensors, networks, cloud computing, and control presents significant opportunities for engineering in general and software engineering in particular. A key goal is to provide certifiable trust in resulting systems, which is a truly formidable challenge for researchers in the field of runtime software assurance.

Over the past 20 years, several research venues (i.e., journals, conferences, and workshops) have emerged in the broad software engineering research community to discuss the design and evolution as well as assurance of SASs.

Mining the rich histories, theories and experiences of fields such as biology, control engineering, and software engineering are worthwhile starting points for as-surance at runtime research. In particular, we need survey papers that investigate models used for design time and runtime assurance techniques in these fields including research on the synergy between them. Moreover, it is useful to relate canonical practical applications to these findings. In a most stimulating 2002 control survey paper Murray *et al.* [MAB[+]02] posit that feedback is a central tool for uncertainty management in modern control. By measuring the operation of a system, comparing it to a *reference* at runtime, and adjusting available control variables, the controller can assure proper operation even in the presence of external disturbances or if its dynamic behavior is not fully known. In software, this reference can be realized with M@RT and evidence for assurance is gathered by checking conformance to the reference model. Murray *et al.* [MAB[+]02] argue that the challenge is to go from the traditional view of control systems as a single process with a single controller, to recognizing control systems as a heterogeneous collection of physical and information systems, with intricate interconnections and interactions [MAB[+]02]. One manifestation of this approach in software engineering is the three levels of runtime control models discussed in Section 3 [TVM[+]13].

The self-adaptive and self-managing systems community has produced a spectrum of runtime models [WMA10] [TVM[+]13] and patterns [RC10b, GH04] with control-centric models [KC03, HDPT04, IBM06, BSG[+]09] at one end and architecture-centric models [BCD97, OGT[+], GCH[+]04, KM07] at the other end. These models come with different attributes and properties that can be exploited for runtime assurance. There is plenty of room for research to compare and evaluate the benefits and synergy of these different runtime model strategies [MKS09, TVM[+]13].

### 4.2 Selected Research Challenges

This section outlines selected open research problems and challenges aligned with the research avenues presented in the previous section. The focus is on the use of M@RT as a basis for developing runtime assurance techniques.

**Runtime Assurance Criteria and Adaptation Properties.** In Section 2.2 we related selected non-functional assurance criteria (e.g., latency) to adaptation properties (e.g., settling time) using quality attributes. One challenge is to extend this characterization of criteria and properties for the target system, controller, and adaptation process. While other approaches may be used to characterize and relate assurance criteria and adaptation properties, the properties are only meaningful if they can actually be measured. Monitoring infrastructure to measure properties is critical for runtime assurance methods. Over the past decade, the SAS community has published numerous papers on various aspects of monitoring. Many of these papers concentrate on the monitoring of raw measures in the managed system but only a limited number of approaches make the information amenable for runtime assurance assessment purposes, including functional requirements monitoring [FF95, FFvLP98, BWS[+]10, DDKM08, MPS08], assumptions monitoring [WSB11, RCBS12], and adaptive monitoring capabilities for changing environmental conditions [RC10a].

**M@RT as a Foundation for Run-Time Assurance.** While M@RT for SAS are increasingly being developed for complex SASs, including reference models [WMA10, VTM+12], few of these models are explicitly designed for runtime assurance. Thus, MART construction for runtime assurance is a key research challenge. The models introduced in Section 3 present good starting points for integrating assurance components into common SAS models. The central challenge for MART construction is to model uncertainty (e.g., environmental disturbances or evolving requirements). Understanding, managing, and leveraging uncertainty is important for delivering SASs with assurance guarantees such as reliability. Ramirez and Cheng [RJC12] have developed a taxonomy of uncertainty commonly faced by SAS, which could be used to facilitate uncertainty modeling and analysis efforts [EKM11, RCBS12]. Fields such as performance engineering and queuing theory have developed advanced models for many different applications. In particular, these fields have developed theories on how to transduce raw measures from a target system into meaningful measures for selected assurance criteria. However, performance constitutes just one dimension of the modeling and assurance problem. Many other quality criteria are applicable to SASs, such as trust, where quantification is rather difficult yet certifiable trust is one of the most important goals for an SAS [Dah10]. Moreover, models are needed to design trade-off analyses schemes for combinations of quality criteria. Models and quality criteria related to governance, compliance, and service-level agreements are of particular importance for service-oriented SASs [BHTV06, TVM+13]. Since M@RT form the foundation of many assurance tasks, the quality of these tasks directly depends on the quality of the models. Defining properties (e.g., accuracy, performance, or safety) for the evaluation of models at runtime is a significant research challenge [TVM+13].

To motivate researchers and practitioners to work on this subject we need compelling reasons for using M@RT for assurance [TVM+13]. A key goal for the SAS assurance research community is to develop exemplars that can be used to evaluate SAS runtime assurance techniques [TVM+13]. Most SAS conferences and workshops regularly call for exemplars but not usually explicitly targeted for SAS runtime assurance. An example of compelling motivation for work in this area is a 20-year science and technology research agenda and outlook for the US Air Force (USAF) [Dah10]. Approximately one third of this agenda is devoted to self-adaptive and autonomous systems with explicit calls for certifiable V&V techniques. V&V is also one of the most promising subfields of assurance where researchers can mine well-established design time models and transition them to runtime. The IBM autonomic computing initiative generated the highly acclaimed MAPE-K [KC03] and ACRA [IBM06] runtime models. The MAPE-K model separates four phases of the feedback loop and thus effectively decomposes the feedback loop assurance problem. The three-layer ACRA hierarchy facilitates integrated assurance reasoning from individually-managed resources at the lowest layer, to managing a collection of resources at the middle layer, to orchestrating an entire system by trading off resource managers at the top layer.

**Run-Time Assurance Methods and Techniques.** For SASs, the boundary between development time and runtime is rapidly disappearing [BG10]. As a result, we need to re-examine the distribution and effectiveness of assurance tasks over the entire life cycle of an SAS. At the same time, we need to determine which models are most appropriate as a foundation for assurance tasks for the different stages of the software life cycle. In particular, we need to investigate whether models that are used for design-time assurance can be effectively used at runtime. In particular, what properties can be guaranteed at development, configuration, or load time as opposed to runtime. While not all assurance tasks can be transitioned to runtime, there is significant opportunity to conduct assurance tasks at runtime thereby making the system more resilient, reliable, responsive, secure, and cost-effective. Regardless of how dynamic a system really is, a substantial part of its assurance will always be done at development time. What (lightweight) design-time techniques can be readily transitioned to runtime? What development-time assurance methods, models, and techniques (i.e., descriptive, prescriptive, constructive and predictive) readily extend to runtime? How do traditional assurance models and methods from domains such as performance, safety, and reliability extend to runtime?

As illustrated in Figure 2, MART play an important role as the abstraction mechanisms required to support every stage of the SAS adaptation process. A key question is what MART techniques are useful for supporting the relevance of runtime monitoring with respect to the assurance criteria. Moreover, to deal with the dynamic nature of functional and non-functional requirements, as well as the execution environment, every component of the adaptation process can also be an adaptive component. Thus, how can M@RT support changes in monitors, analyzers, planners and executors according to changes in functional and non-functional requirements? In the realm of control system engineering, changing the controller is referred to as *adaptive control* [AW94]. Another important avenue of research is how to characterize runtime assurance techniques according to the different levels of dynamics in SASs (i.e., changes in requirements, relevant context, adaptation mechanisms, and the target system itself).

Assurance obligations vary from one application domain to another. For example, the area of safety-critical systems has developed specialized assurance criteria and models—albeit mostly design-time techniques (e.g., ISO26262 for automotive subsystems,[7] and numerous safety standards set by the International Electrotechnical Commission).[8] The service-oriented architecture (SOA) community has developed SOA governance models—a combination of design time and runtime models—for assurance tasks for service-oriented systems on SOA platforms [SMB$^+$09]. Thus, it is useful for researchers to classify runtime assurance criteria, models, and techniques according to their applicability to different domains and applications (e.g., application-independent, domain-dependent, mission-critical systems, embedded systems, real-time systems, etc.). Run-time assurance techniques can also be classified according to different types of runtime changes (e.g., dynamic context, changing requirements, or evolving models).

With the increasing use of computing-based systems for delivering critical societal services that demand long-running or even continuous operation (e.g., telecommunication, power grids, financial systems, etc.), even in the face of adversity, adaptation and runtime evolution [MV14] is a necessity, not a luxury. Even with meaningful reactions to changes, the triggered SAS adaptation should preserve selected core properties, thus posing a need for incremental and compositional assurance for SASs. An enabling step, in this direction, is to split functional and non-functional requirements into sub-requirements associated with single services and components of the system. The idea is to decompose the requirement specification into properties associated with the behavior of small parts of the system. Thus, it becomes possible to check these properties locally and to deduce from local checks whether the system satisfies the overall specification. By decomposing the assurance task in such a way, it may not be necessary to build a complete model of the system and thus the combinatorial state explosion problem is mitigated. The main challenge of this approach is that local properties are typically not preserved at the global level because of dependencies among the aggregate subparts of the system. Another approach to decomposing the assurance problem is to separate the verification of the functional properties from the verification of adaptation properties. Zhang *et al.* [ZGC09] developed AMOEBA, a modular verification approach for SASs where the functional properties are specified in terms of LTL and the adaptation properties are specified in terms of A-LTL [ZC06b]. With this separation of concerns, AMOEBA uses an assume/guarantee approach [JT96] to perform incremental model checking of both types of properties. AMOEBA-RT is an extension that monitors the adaptation properties at runtime based on state-based models of the adaptive logic [GCZ08].

---

[7] http://www.iso.org/
[8] http://www.iec.ch/

As another example of assurance for the adaptation process, suppose *settling time* (i.e., the time required for the adaptation mechanism to take the target system to the desired state) has been defined as a performance-oriented assurance concern for a particular adaptive system. As such, the assurance mechanisms must keep track of the time the adaptation mechanism is taking to complete the adaptation process—generally goals must be reached within a suitable time interval. An extremely long adaptation process could render the system to be useless or even detrimental to the system's overall safety. The desired thresholds, monitoring conditions, and entities to be monitored can be specified using M@RT, such as goal-based models [WSB$^+$09] or contextual RDF graphs [VMT11a, VMM$^+$11].

## 5 Characterizing Assurance Methods

Researchers from communities related to the engineering of SASs have contributed a spectrum of approaches to the assessment of adaptive software. Rather than producing a comprehensive and systematic literature review of the state of the art, the goal of this section is to provide an overview of how M@RT have been used as runtime assurance enablers in selected domains. This characterization of assurance approaches provides a starting point upon which researchers can build to address the research challenges posed by model-based runtime assurance of SASs.

### 5.1 Classifying Assurance Methods According to Techniques

This section presents and classifies selected existing approaches for runtime assurance of SASs according to the techniques and methods used for their realization.

**Goal-Oriented Approaches.** A first step towards assuring software systems is the articulation of assurance criteria. This task can be complex for functional requirements because it requires a deep understanding of the application domain. Nguyen *et al.* [NPT[+]09] argue that goal-oriented techniques are effective for deriving assurance criteria from functional requirements specifications. At development time (or requirements negotiation time), goal models can be used to specify stakeholder expectations for SASs, and the decision criteria for acceptable system behavior can be derived from these models. Moreover, goals, and especially high-level goals, have been recognized as more stable (i.e., less volatile) than specific system requirements [vLDL98]. Thus, high-level goals provide suitable candidate assurance criteria in highly dynamic systems. Qureshi *et al.* [QJP11, QLP11, QP10] rely on this assumption in their work on continuous requirements engineering. They represent functional behavior in terms of high-level goals (i.e., functional goals) that are decomposed into sub-goals. Alternative decompositions are qualified by quality criteria, user preferences, and context that contribute positively or negatively to their ranking. To ensure the expected behavior, the system must select the most appropriate goal decomposition path.

The effectiveness of the assurance of SASs at runtime is highly dependent on the changing conditions of the execution environment that can affect not only the target system, but also the adaptation mechanism and monitoring infrastructure. Ramirez and Cheng proposed an approach to manage changes in monitoring conditions according to environmental situations at runtime [RC11]. They specify requirements goal models using the RELAX language [WSB[+]09]. Recently, AutoRELAX has been developed to automatically add RELAX operators to goal models to handle uncertainty in the environment while minimizing the number of reconfiguration adaptations [FDC14a]. In a similar approach, Pasquale *et al.* [BPS10, PS11] developed FLAGS, a KAOS goal modeling framework that introduces the concept of a fuzzy goal whose satisfaction can be evaluated through fuzzy logic functions. Both goal-modeling approaches use fuzzy logic-based functions to add flexibility to the satisfaction criteria of goals in a goal-oriented model. In contrast to RELAX, however, FLAGS does not focus on identifying sources of uncertainty, but focuses rather on evaluating the degree to which a goal is satisfied. Goal-based models can be transitioned from design time to runtime to track changes in SAS requirements at runtime. Morandini *et al.* have investigated the life-cycle of goals at runtime [MPP09]. Souza *et al.* [SSLRM11] have developed a system, *Zanshin*, a requirements monitoring framework based on multiple feedback loops to monitor awareness requirements and progress towards adaptation objectives at runtime [ASaP13].

**Automatic Test Case Generation- Based Methods.** The complexity of system structure and behavior is growing exponentially, coupled with the comparable volume of possible scenarios and combinations of environmental conditions to be handled by an SAS. As such, successful strategies for automatic test case generation used for non-SAS application areas are being leveraged and explored for SAS testing. For example, given that multi-agent based software systems expose high levels of runtime dynamism, applicable testing techniques for these types of systems can be leveraged to assess SASs using M@RT [NPB[+]09]. An important challenge in the validation of SASs at runtime using direct-testing techniques is the generation of test cases that are relevant to the system's current execution context and goals. As a means to evaluate system performance, Nguyen *et al.* [NPT+09] use evolutionary testing techniques to automatically generate test cases based on quality functions. Quality functions are associated with stakeholder expectations of the behavior of an autonomous system which are expressed as goal-oriented requirements. (e.g., the quality function associated with the goal of a cleaning agent to maintain its battery can be a minimum battery level to be satisfied). This approach allows the automatic generation of test cases with increasing difficulty levels, guided by a fitness function associated to the quality of interest (e.g., a function inversely proportional to the total power consumption of the system throughout its lifetime). A complementary approach is taken by Fredericks *et al.* [FRC13b] where an SAS is exposed to a wide range of adverse environmental conditions that are used to generate SAS execution traces as the system adapts and reconfigures to handle the adverse conditions. These traces can then be analyzed for unexpected and/or unwanted behavior, both in the functional and in the adaptive logic. EvoSuite [FA11] is a framework that implements an evolutionary algorithm to generate test suites that consider a single coverage criterion, for instance the introduction of artificial defects into a program. Finally, a MAPE-T loop [FRC13a] has been proposed to provide a framework for monitoring the applicability and utility of test cases for an SAS as it undergoes environmental changes and reconfiguration. A set of research challenges were posed as part of the proposed framework, including explicit reference to the importance and use of M@RT. Veritas [FDC14b] is a recent realization of the MAPE-T loop that adapts test cases to ensure testing relevancy as an SAS reconfigures to handle changing environmental conditions.

**Model Checking.** Model checking [CGP01, PPS09] was proposed in the 1980s independently by Clarke and Emerson [CE82], and Quielle and Sifakis [QS82]. It assumes an available mathematical model of a system and a property to check against the model expressed in a formal logic, such as Linear Temporal Logic (LTL) [Pnu81] or Computational Tree Logic (CTL) [BAMP81]. The goal of model checking is to use an algorithmic approach to check the consistency between the given model and the property specification. Model checking has been used extensively to verify hardware [BLPV95] and software systems [CGP02] in many application domains to assure desired properties. Model checking at runtime is a key strategy to verify SASs based on runtime models. Weyns *et al.* surveyed formal methods in self-adaptive systems [WIdlIA12]. They showed that there are no standard tools for formal modeling and verification of self-adaptive systems. According to their survey, however, 40% of the surveyed studies use tools for formal modeling or verification, and 30% of those studies use model checking tools.

A number of model checking techniques have been used to analyze various properties of SASs. Baresi *et al.* used model checking to check whether an architecture is a refinement of another one [BHTV06]. Specifically, they defined refinement relationships between abstract and concrete styles. The defined refinement criteria guarantee both semantic correctness and platform consistency. In another approach, Abeywickrama and Zambonelli proposed to model check goal-oriented requirements for SASs [AZ12].

Cámara and de Lemos used probabilistic model checking to verify resilience properties of SASs, with the goal of verifying whether the self-adaptive system is able to maintain trustworthy service delivery in spite of changes in its environment [CdL12]. In architecture-based domains, Pelliccione *et al.* applied model checking at the software architecture level to verify properties of the system, its components, and the interactions among components [PIM09, PTBP08]. Filieri *et al.* have developed a runtime probabilistic model checking technique to detect harmful reconfigurations. To deal with unplanned adaptations, Inverardi *et al.* proposed a theoretical assume-guarantee framework to define under which conditions to perform adaptation by still preserving the desired invariants [IPT09]. Zhang and Cheng developed AMOEBA [ZGC09], a modular model checker to separately verify SAS functional properties in terms of LTL and the adaptive logic in terms of A-LTL (adapt-LTL). AMOEBA-RT [GCZ08] verifies runtime properties of SAS properties. Model checking has also been applied in the domain of agent-based systems, for instance to assure adaptability to unforeseen conditions, behavioral properties, and performance [Gor01]. Finally, Murata used Petri Nets to enable the analysis of properties, such as the reachability of a certain state or deadlock-freeness [Mur89]. Some of these analysis methods have been extended to enhanced versions of Petri Nets, such as Colored Petri Nets [Jen03] and applied to check properties such as performance [Wel02] or safety [CHC96].

**Rule-Based Analysis and Verification.** Several approaches based on formal methods, especially graph-based formalisms, have been proposed to leverage rule-based analysis and verification of software properties. In particular, Becker and Giese proposed a graph-transformation based approach to model SASs at a high-level of abstraction. Their approach considers different level of abstractions according to the three-layer SAS reference architecture proposed by Kramer and Magee [KM07]. In their approach, Becker and Giese check the correctness of the modeled SAS using simulation and invariant-checking techniques. Invariant checking is mainly used to verify that a given set of graph transformations will never reach a forbidden state. This verification process exposes a linear complexity on the number of rules and properties to be checked [BBG$^+$06]. In another approach, Giese *et al.* used triple graph grammars as a formal semantics for specifying models, their relation, and transformations. These models can be used as a basis for analyzing the fulfillment of desired properties [GHL10]. In the self-healing domain, Bucchiarone *et al.* proposed an approach to model and verify self-repairing system architectures [BPVR09]. In their approach, dynamic software architectures are formalized as typed hyper-graph grammars. This formalization enables verification of correctness and completeness of self-repairing systems. This approach was extended later by Ehrig *et al.* [EER$^+$10] to model self-healing systems using algebraic graph transformations and graph grammars enriched with graph constraints. This extension enables formal modeling of consistency and operational properties. In the quality-driven component-based software engineering domain, Tamura *et al.* [TCCD12, Tam12] formalized models for component-based structures and reconfiguration rules using typed and attributed graph transformation systems to preserve QoS contracts. Based on this formalization, they provide a means for formal analysis and verification of self-adaptation properties, both at design time and runtime by integrating the Attributed Graph Grammar (AGG) system in their framework.

**Synthesis.** Another interesting avenue of research is to use synthesis techniques for assuring SASs. The goal of these techniques is to generate the "correct" assembly code for the (pre-selected and pre-acquired) components that constitute the specified system, in such a way that it is possible to guarantee that the system exhibits the specified interactions only. Inverardi *et al.* [IST11] proposed a synthesis-based approach for networking. This approach considers application-layer connectors by referring to two conceptually distinct notions of connector: *coordinator* and *mediator*. The former is used when the networked systems to be connected are already able to communicate but they need to be specifically coordinated to reach their goal(s). The latter goes a step further by representing a solution for both achieving correct coordination and enabling communication between highly heterogeneous networked systems. This work has been extended to also handle non-functional properties [DMIS13]. La Manna *et al.* [PGGB13] proposed an approach for reasoning about safeness of dynamic updates based on specification changes.

**Semantic Web.** A key challenge for establishing runtime assurance of SASs is the preservation of the relevance of runtime monitoring infrastructures with respect to assurance criteria and the system's execution environment. Specifically, monitoring strategies and infrastructures must adapt themselves dynamically. Models at runtime are also required to support self-adaptation of context management infrastructures (i.e., the third level of dynamics in SASs that was presented in Sect. 3.1). To manage context dynamically, the explicit mapping between assurance concerns and relevant context must be complemented with an explicit mapping between relevant context and infrastructure elements of the monitoring infrastructure. In this way, whenever changes in assurance criteria or relevant context occur, the dynamic adaptation of a representation of the monitoring strategy will trigger the adaptation of context sensors, context providers, and context monitors accordingly. Ramirez and Cheng [RCM10] used a goal-based approach to adapt the monitoring infrastructure to support the changing execution context for an SAS. Resource description framework (RDF) graphs, from semantic web, are good candidates to be used as effective M@RT in the assessment of SASs. Models at runtime in the form of RDF graphs can be exploited to represent relevant context, monitoring strategies, system requirements including assurance criteria, as well as to support changes in context management strategies at runtime. Ontologies and semantic-web based rules, defined according to the application domain, provide the means required to infer changes in the monitoring infrastructure according to changes in requirements, assurance criteria or context [VMT11a, Vil13].

### 5.2  Classifying Assurance Methods According to Non-Functional Criteria

In this subsection, we classify surveyed runtime assurance approaches according to the non-functional requirements they address as assurance criteria.

**Safety.** For systems that are self-adaptive or even self-organizing, the application of traditional safety assurance approaches is currently infeasible. This obstacle is mostly due to the fact that these approaches rely heavily on a complete understanding of the system and its environment, which is difficult to attain for adaptive systems and as of yet impossible for open systems. Open systems, in contrast to self-adaptive systems that are generally closed systems, do not use measured outputs to determine control inputs required to adjust their behavior [HDPT04]. Therefore, open systems necessarily require a complete and accurate model of the system and its environment from which the control input must be derived. These models are generally impractical given that they must be robust to changes in the system and its environment and use no feedback mechanism to adjust themselves. A general solution is to shift parts of the safety assurance measures into runtime when all required information about the current state of the application can be obtained. Rushby [Rus07] developed a strategy where development-time analysis techniques for certification are used at runtime, but the actual certification is performed as needed just-in-time. Based on this work, he later coined the notion of runtime certification [Rus08], using runtime verification techniques to partially perform certification at runtime. Following the same core idea of shifting portions of the assurance measures into runtime, Schneider *et al.* [ST13] introduced the concept of conditional safety certificates (ConSerts). ConSerts are predefined modular safety certificates that have a runtime representation to enable dynamic evaluations in the context of open adaptive systems. Some initial ideas concerning the extension of ConSerts regarding other certifiable non-functional properties such as security have also been published [SBT11]. Priesterjahn and Tichy [PT09] proposed a different approach based on the application of hazard analysis techniques during runtime. This approach is closely related to their previous work where they introduced a development-time hazard analysis approach for analyzing all configurations that a self-adaptive system can reach during runtime [GT06]. A corresponding extension also considers the time between the detection of a failure and its reconfiguration [PSWTH11].

**Performance.** Regression models and queuing network models (QNM) are M@RT commonly used to reason about performance-based assurance properties relating to response time, throughput, or utilization. For example, Hellerstein *et al.* [HDPT04] and Lu *et al.* [LAL$^+$03] described dynamic regression models in the context of autonomic computing and self-optimization. Menascé and Bennani [MB03] used QNM as predictive models for avoiding bottleneck saturation and for online capacity sizing. Ghanbari *et al.* [GSLI11] used dynamically tuned layered queuing models, which are software specific versions of QNMs, for online performance problem determination and mitigation in cloud computing. More recently, Barna *et al.* [BLG11] reported performance load and stress testing methods on online tuned runtime performance models.

**Reliability and Availability.** Run-time assurance methods for reliability and availability properties use discrete time Markov chains that are synchronized with the system and its usage profile. For example, service-based systems built using the QoSMOS (QoS Management and Optimization of Service-based systems) framework [CGK$^+$11] translate high-level QoS requirements specified by their administrators into probabilistic temporal logic formulae that are then formally and automatically analyzed to identify and enforce optimal system configurations. The QoSMOS self-adaptation mechanism can handle reliability and performance-related QoS requirements. QoSMOS [FGT11, MG10] uses the KAMI approach [EGMT09] to keep the model, including its parameters, and the system consistent; it uses probabilistic model checking at runtime to evaluate whether the system satisfies the current reliability requirements.

**Security.** Security considerations revolve around self-protection goals of an SAS, including confidentiality, integrity, authenticity, and authorization [BCdL11, KHW$^+$01]. Run-time assurance of these goals is important in SASs since adaptation may produce emergent behavior that violates one or more other critical system properties. In particular, security assurance must be achieved without compromising system goals unrelated to security [RZN05, HMPB00]. For example, security considerations, such as confidentiality may conflict with availability goals. While the former, confidentiality, aims to protect the information in the system from unauthorized access, the latter, availability, is intended to ensure access to the system and the information a user is authorized to access. One way of counteracting an intrusion is by limiting access to the parts of the system that are affected by an attack. This approach clearly can have negative impact on availability. It is therefore important that, within an SAS, any remedial interventions invoked to preserve security goals also preserve the system properties not related to security. Achieving this balance requires decisions to be made at runtime based on evidence regarding the satisfaction of security goals obtained from analyzing the system and its environment, including user behavior.

Run-time security of an SAS involves not only protecting the target system, but it also means that the adaptation process and the policies governing the adaptation are protected from malicious attacks (e.g., preventing attackers from hijacking its adaptation mechanisms and policies) [Ais03, BJY11, OMH$^+$11]. Adaptation methods, data, policies and certificates must be properly protected to ensure confidentiality, authenticity, and trusted communication of the entire adaptation process and its drivers. The components of every MAPE-MART loop depicted in Figure 2 must also be protected accordingly.

While an SAS is expected to make its adaptation decisions autonomously, a key question is how and how much to empower users with privacy and data security control (e.g., when user context is involved in adaptation decisions). The Surprise [MTVM12] approach (i) allows users to configure access permissions to their sensitive personal information to third parties, selectively and with different levels of granularity; (ii) supports changes in these configurations at runtime to add or remove third parties or permissions, and (iii) realizes partial encryption to share non-sensitive data with third parties who have not been explicitly authorized access, while protecting user identity. The Surprise approach is an exemplar of the application of M@RT to the preservation of privacy and security policies in user-driven SASs.

Security assurance, like other assurance goals at runtime, relies on the definition of high-level policies that must be preserved during adaptation. To achieve this security assurance, the Self-Adaptive Authorization Framework (SAAF) uses a feedback loop that continuously monitors the decisions made by the system's authorization pro-cess [BCdL11] . The knowledge gained is used to adjust the authorization policy at runtime, making it more restrictive to constrain user behavior or loosening it to endorse users. Dynamic conflict resolution is particularly important in the context of security assurance but many existing approaches, e.g. [HMPB00], resolve conflicts using priority levels assigned at design time. Instead, the ATNAC (Adaptive Trust Negotiation and Access Control) framework [RZN05] allows access control policies to be dynamically adjusted depending on a set of trust-associated attributes observed at runtime. Formal methods have also been used successfully in this context. For example, the Willow Architecture [KHW$^+$01], a dynamic reconfiguration framework for critical distributed systems, enables systems to continue working with reduced functionality while under a security attack. The use of formal methods enables autonomous handling of conflicts at runtime during reconfiguration.

**Usability.** In applications with adaptive user interfaces, it is often impossible to test each adaptation state with real users. Therefore, automated usability evaluation of such user interfaces often relies on models of the user or user interactions to evaluate states of user interfaces automatically [IH01]. Quade *et al.* [QBL$^+$11] introduced an approach that evaluates the usability of the current state of a user interface using M@RT. The evaluation is based on a simulation of user interactions based on the model of the user interface and a model of the user. Having these techniques available at runtime enables a more detailed modeling of the user as the model can be checked against data from the actual user interaction.

## 6   Compelling Applications for Models at Runtime

This section introduces application exemplars for which M@RT play a major role in the assurance of functional and non-functional assurance criteria. The goal of this section is to provide a catalogue of "killer applications" useful to motivate case studies on the assurance of SASs where M@RTare used as a foundation.

**Kaleidoscope.** Kaleidoscope [9]is a multi-channel multimedia video streaming and video on demand system. Imagine an Olympics game or a football match where millions of users are simultaneously streaming, watching and querying videos about the event. The Kaleidoscope application aims to provide/share best quality video for its users. As such, Kaleidoscope must act as a proxy server that is used to store and forward multimedia content to user devices. A device can be a notebook, a smartphone, or a personal digital assistant (PDA). Kaleidoscope must detect both the video source and the user target device. Kaleidoscope must adapt at runtime from one configuration variant to another in order to provide the best quality video to users concurrently and reliably. The broadcast is fetched from a video source via TV cable (e.g., TV broadcast) or either wired or wireless (e.g., Webcast) Internet connection.

Latency and capacity (i.e., bandwidth) are important assurance criteria in Kaleidoscope since high-quality video streaming is a major functional requirement. To guarantee functional requirements under the desired quality conditions, Kaleidoscope must adapt itself by reconfiguring its network and software architecture to minimize latency and maximize capacity. In this scenario, M@RT are useful for a variety of purposes. For example, predictive models can be used to anticipate latency and required capacity in the near future to perform preventive adaptations and thus avoid the violation of the desired qualities. Another example is the use of runtime formal models such as those exploited in rule-based analysis and verification to guarantee the reliable re-configuration of the system.

---

[9] http://www.savinetwork.ca

**Autonomous Vehicle Service.** Google driverless cars are now licensed in California, Florida and Nevada.[10] Google engineers and scientists achieved this amazing feat in a short five years after DARPA formulated the Great and Urban Challenges on autonomic cars.[11]

It is speculated that driverless cars could come from and go to parking lots, or deliver packages. In a carpooling scenario, autonomous vehicles booked by users could serve the user at a specific time and destination. Best routes will be planned intelligently based on current context information such as traffic conditions and weather. Ordering, booking, and payment will be performed via smartphone applications. Elderly people will become mobile again, as they will be have greater access to services using an autonomic vehicle.

Increasingly, cars are being equipped with intelligent driver assistance for anticipating potential hazards early and avoiding collisions. Intelligent, yet safe autonomous driving software systems require effective methods to ensure their required qualities. Even though the functions of these vehicles are perceived as "intelligent", they typically rely on standard algorithms from sensor fusion, context management, and control theory. In particular, these systems require special attention to context management infrastructures to guarantee the reliability of sensors and monitors. Autonomous vehicle software use models at several levels, especially for understanding relevant context situations: models are required to represent entities that affect the behavior of the car, to specify quality of sensors, and to model context uncertainty. Given the dynamic nature of context information, these models must be available and manageable at runtime. Another category of important models are those that specify typical vehicle behavior used to understand unusual behavioral patterns.

Models for autonomous vehicle software are typically developed implicitly and coded manually into the running system. In order to rigorously address the assured behavior of these systems, these models need to be managed explicitly and rigorously throughout the software life cycle, including at runtime.

**Autonomous Agricultural Operations.** *Precision agriculture*[12] is an approach to realize a comprehensive farming management concept. One of the main issues addressed by precision agriculture is the optimization of the productivity and efficiency when operating on the field, by tailoring soil and crop management to match the conditions at each location. This level of customization can be achieved through the use of different information sources such as GPS, satellite imagery, and IT systems. More recently, efforts have been underway to further improve productivity and efficiency by increasing the amount of automation on the field to the point of autonomous operation. Examples are harvesting fleets comprising several harvesters but only one is operated by a human, autonomous tractors that pick up the crop from the harvesters, and tractor implement automation (TIA) where tractors are controlled by implements to execute implement-specific tasks. These application scenarios have in common that different vehicles or machines are combined on the field in order to fulfill (partially) autonomous tasks. The assurance and certification of important properties, such as safety and security are clearly critical in this context. Furthermore, traditional assurance techniques are not applicable without significant modifications. A first step to this problem is to shift parts of the assurance measures into runtime. This strategy can be achieved by means of suitable M@RT and corresponding management facilities integrated into these systems.

---

[10] http://www.forbes.com/sites/ptc/2013/11/06/why-google-and-others-see-a-future-with-driverless-cars/print/
[11] http://www.tartanracing.org/challenge.html
[12] https://www.ispag.org

**Ambient Assisted Living.** The number and capabilities of devices available at home are growing steadily. *Ambient Assisted Living (AAL)* is intended to use these technologies to assist users with disabilities in their daily tasks, such as monitoring health conditions and detecting emergency situations.[13] Software applications in this domain are not only critical, but also highly dynamic. On the one hand, human lives can be compromised. On the other hand, every home is different and can contain different devices that could be leveraged by AAL services. New generations of devices are produced on a regular basis requiring AAL services to evolve continuously to keep up to date with new technical developments. Moreover, similar devices produced by different vendors may differ considerably in their capabilities and interfaces. Nevertheless AAL systems must be able to use these devices as soon as they become available at the user's home in an effective and safe manner.

In addition to variations in devices, users of AAL systems are subject to considerable variation. An AAL service must deal with an arbitrary number of people living at the same home, their disabilities and capabilities, and their current environmental conditions. Therefore, the system is required to adapt itself according to current users and their environment. Moreover, these systems must be sufficiently flexible to support future extensions, such as the integration of new sensors or actuators for new applications. Most importantly, these adaptations must be performed seamlessly and reliably to guarantee user safety.

To deal with these complex dynamics, AAL software requires M@RT to reason about users and their context in order to correctly and safely deliver services. Moreover, it is important to maintain a causal connection between these models and both the target systems and adaptation mechanisms. Given the potential risks to human lives, assurance is a major concern that must be guaranteed to prevent hazardous operation before, during, and after adaptation [ZC06a, VMT$^+$11b]. M@RT can be essential in the management of AAL software for capturing the environment, monitoring the user interaction, and reasoning about possible adaptive behavior and their impact.

---

[13] http://www.aal-europe.eu

**The Guardian Angels Project.** In the context of AAL, the "Guardian Angels for a Smarter Planet" project[14] is a good example to illustrate the potential benefits from using M@RT to address SAS assurance. The following details are based on information from the Publications Office of the European Union[15]:

The overarching objective of the Guardian Angels Flagship Initiative is to provide information and communication technologies to assist people in all stages of life. Guardian Angels are envisioned as personal assistants. They are intelligent (thinking), autonomous systems (or even systems-of-systems) featuring sensing, computation, and communication, and delivering features and characteristics that go well beyond human capabilities. It is intended that these systems will provide assistance from infancy through old age. A key feature of these Guardian Angels will be their zero power requirements as they will scavenge for energy. Example services include individual health support tools, local monitoring of ambient conditions for dangers, and emotional applications. Scientific challenges for supporting their research challenges include energy-efficient computing and communication; low-power sensing, bio-inspired energy scavenging, and zero-power human-machine interfaces.

These devices, by their very nature, will need to be adaptive in terms of functional and non-functional properties. In addition, they will be used in critical situations that require high levels of dependability and hence the highest levels of safety assurance.[16] The development of M@RT can support runtime decision making and certification for this important and innovative application area.

## 7 Conclusions

This chapter presented a research agenda for assurance at runtime with M@RT as a foundation. It grew out of stimulating discussions among the participants of the 2011 Schloss Dagstuhl Seminar on Models@run.time. In particular, we report on the findings of the breakout group Assurance@run.time as well as online discussions among the authors over the past two years while writing this chapter.

In an instrumented, interconnected and intelligent world, self-adaptive software systems proliferate. A key goal is to provide assurance at runtime when such systems adapt at runtime due to changes in their execution environment or their requirements. Traditionally software engineering, as opposed to control engineering, has concentrated on design-time assurance. Thus, a key challenge for the software engineering community is to develop runtime assurance techniques for self-adaptive systems that provide high performance, high confidence, and reconfigurable operation in the presence of uncertainties. One of the most promising avenues of research in this area is to use M@RT

---

[14] http://www.ga-project.eu

[15] Publications Office of the European Union: FET Flagship Pilots, Community Research and Development Information Service (CORDIS), http://cordis.europa.eu/fp7/ict/programme/fet/flagship/6pilots_en.html, 2012.

[16] http://www.ga-project.eu/science/software

as a foundation for developing runtime assurance techniques. Of all the subfields of assurance, V&V has probably made the most progress in transitioning design time models and techniques to runtime. While not all design-time assurance tasks can be transitioned to runtime, a significant opportunity exists to conduct assurance tasks at runtime, thereby making the overall SAS more resilient, reliable, responsive, secure, and cost-effective. One of the most formidable challenges for researchers in the field of runtime software assurance is to investigate techniques that guarantee certifiable trust for highly-adaptive systems.

This research agenda on runtime assurance techniques provides excellent starting points for research communities dealing with SASs, including Models@runtime, Run-time V&V, Requirements engineering@runtime, SEAMS, SASO (International Conference on Self-Adaptive and Self-Organizing Systems), and ICAC (International Conference on Autonomic Computing). Given the increasing use of SAS for high-assurance application domains, such as intelligent vehicles, power grid management, telecommunication infrastructure, financial systems, healthcare management systems, etc., it is paramount that these communities and related communities work together to address the assurance of SASs. M@RT is a key enabling technology to accelerate progress in this area.